# Resolving the relaxation complexity of vitrimers: time-temperature superpositions of a time-temperature *non-equivalent* system

Paolo Edera [a,*], Selene Chappuis[a], Michel Cloitre[a], Francois Tournilhac[a]

[a] Molecular, Macromolecular Chemistry, and Materials, CNRS, UMR 7167, ESPCI−Paris, PSL Research University, 10 Rue Vauquelin, 75005 Paris, France.

Email: paolo.edera@espci.fr

ABSTRACT:

Vitrimers are polymer networks that, thanks to covalent bond exchange, combine desirable properties of thermoplastic and thermosets, such as flowability and insolubility. For this reason, vitrimers are considered to be good candidates for a number of innovative applications from self-healing soft robots to hard reprocessable materials. All these applications are related to the unusual thermomechanical behavior of vitrimers, consequence of the non-trivial interplay between the polymer network dynamics and the thermally activated chemical link exchange. Here we use solid-state rheology to investigate the properties of a recently developed epoxy-based vitrimer. The rheological analysis demonstrates that the mechanical spectrum is composed of two relaxation processes with distinct activation energies which are associated with glass dynamics and covalent bond exchange, respectively. This makes the material thermo-rheologically complex and time temperature equivalence does not apply. Nonetheless, thanks to mechanical spectral analysis in a wide range of stiffness, time and temperature, we are able to depict the time-temperature-relaxation landscape in an enough precise way to account for the two dynamical processes and recombine them to predict the mechanical moduli in a wide (virtually unlimited) interval of frequencies, from low temperatures (close to room temperature) to high temperatures (above the Tg).

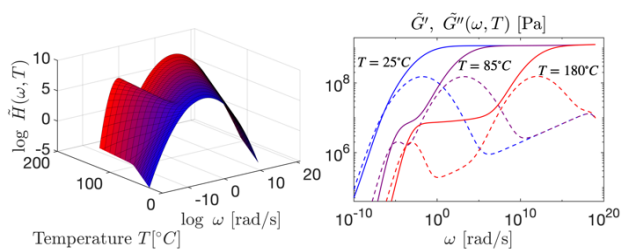

# 1. Introduction

By constitution, vitrimers are polymeric networks with thermally exchangeable links, which makes them able to flow when heated while remaining essentially insoluble, even in a good solvent where they swell without dissolving [1]. They thus share the cohesiveness of thermosets and the reprocessability of thermoplastics. This unusual combination of properties makes them recognizable as a third class of polymer materials and desirable in applications. Materials for soft robotics, which are rapidly damaged when they encounter sharp objects, could benefit from the self-healing ability of vitrimers without adverse effect on creep resistance [2]. At the opposite of the mechanical spectrum, hard polymer matrices for fibre reinforced composites are difficult to recycle because implementation methods such as resin transfer molding (RTM) are ideally suited for thermosets [3], which are also those offering the best thermomechanical performance. In aerospace composites the reprocessability, reparability and recyclability enabled by the chemical constitution of vitrimers would be definitely an advantage.

In a recent paper, it has been shown that using conventional resins, a hardener easily available commercially, and a classical manufacturing technique, the glass transition temperature of epoxy-based vitrimers can be increased to more than 200°C while the curing at any temperature can be represented synoptically at using a TTT (time, temperature, transformation) diagram [4]. The objective of this present paper is to present the time/temperature relationship of viscoelastic properties of hard cured vitrimers in a synoptic form, being understood that to describe those vitrimers in all their complexity, it is necessary to cover more than 4 orders of magnitude of complex modulus and 20 orders of magnitude in time.

In vitrimers, as in other polymeric systems, the principle of time-temperature superposition (TTS) has been used to access a wide frequency domain [5, 6, 7]. But in some cases, the TTS principle fails [8, 9, 10, 11, 12, 13, 14, 15]. In these cases, a strategy that has been used is to collapse the viscoelastic moduli successively at high and low frequencies to obtain two different activation energies. Whenever the temperature window of interest includes the glass temperature transition ($T_g$), two different physical processes come into play to produce the mechanical response of the material: the dynamics of the polymer chains composing the network and that of the reversible covalent links. The two processes have in general different activation energies, and thus different dependence on temperature, which makes them thermo-rheologically complex, so that the time temperature equivalence does not hold[16]. On the other hand, these ingredients not necessarily hinder the possibility to collapse the measures performed at different temperatures in a single master curve, as it has been reported in experiments [17] and simulations [18]. To summarize, even if time temperature equivalence does not hold due to the presence of different processes with different temperature dependence, the time temperature superposition does not necessarily fail, meaning that it is possible to superimpose experiments at different temperatures using shift factors on the frequencies. Intrigued by these observations, and interested in assessing the validity of time temperature superposition in vitrimers, we present here an analysis based on the relaxation time distribution H($\omega$) [19], which gives evidence for two separated families of relaxation times, each one with its own temperature dependence. As we will show, this analysis allows us to reconstruct the viscoelastic moduli over an impressive range of temperatures and frequencies, despite the failure of time temperature equivalence. The methodology is demonstrated for a recently developed system [20] whose mechanical properties are studied both below and above $T_g$. The so built viscoelastic moduli can be modelled with the sticky Rouse model [11, 21, 22], giving access to microscopic properties, such as the average size of the permanently connected objects (number of monomers), lifetime of dynamic bonds, and number of dynamic bonds per permanently connected object.

## 2. Experimental

### 2.1 Materials chemistry and elementary topology

The chemical workflow is shown in Figure 1. Glutaric anhydride (GA, $n$ mol.) is reacted with bisphenol A diglycidyle ether (DGEBA, $nq-h$ mol.) in the presence of bisphenol A *bis*(2,3-dihydroxypropyl) ether (DHEBA, $h$ mol.) and Zn (II) acetylacetonate (Zn(acac)$_2$, 10 mol% with respect to epoxy functions) as described in reference [20]. This stoichiometry is formally equivalent to reacting $n$ mol of GA, $nq$ mol. of DGEBA and $h$ mol of water. Adding a defined quantity of DHEBA permits to control the number of ester, ether and hydroxide functions throughout the network and keep the glass transition temperature basically unchanged. Molar quantities actually used in this work are reported in Table 1, giving rise to a material with glass transition of about $T_\alpha \approx 100°C$. A picture of the network building blocks thereby generated, including the average number of exchangeable and non-exchangeable links is shown in Figure 1b-c, while the characteristic dimensions thereof are reported in Table 2.

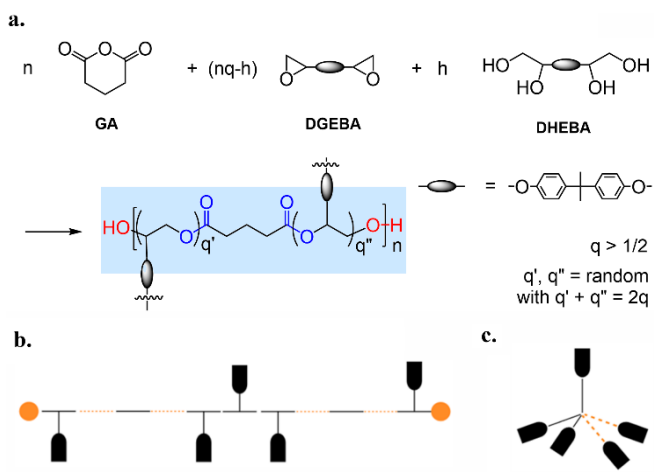

**Fig. 1**. Network synthesis from glutaric anhydride (GA), bisphenol A diglycidyl ether (DGEBA) and bisphenol A *bis*(2,3-dihydroxypropyl) ether (DHEBA). **a.** Stoichiometry of the reaction, **b.** Illustration in 2D of an average resulting molecule when using $n = 2$ and $q = 5/4$. **c.** Illustration of the same, seen as a 3D network node; black lines figure permanent links, broken orange lines figure exchangeable links.

Table 1: Stoichiometry of the reaction reduced to 1 mol of diepoxide.

|  | General formula | This work |
|---|---|---|
| GA | $n$ | 1 |
| DHEBA | $h$ | ¼ |
| DGEBA | $nq - h$ | 1 |
| Zn (II) (acac)$_2$ | 5 to 10 mol% | 0.05 |

Table 2: Characteristic sizes of the elements forming the network nodes.

|  | General formula | This work |
|---|---|---|
| Number of diacyl linkers between two OH | $n/2h$ | 2 |
| Number of branches between two OH | $2nq/h + 1$ | 5 |

*2.2 Rheological measurements*

Torsional rheology is performed using an Anton Paar 702 multidrive analyzer. The specimen dimensions are: $45 \times 10 \times 4$ mm$^3$. The normal force is imposed to be -1N, corresponding to a normal tensile stress of 25 kPa. Small amplitude oscillatory shear (SAOS) experiments are performed in the [0.01–100] rad/s angular frequency range. Data affected by large inertia corrections are discarded, as described in SI. The linear regime at 1 Hz and at different temperatures (40, 90, 180 ˚C), is determined to be (0.01%, >1%, >1%) respectively (see SI Fig. S2). The temperature is controlled by the oven (Anton Paar CTD 600), each time it is changed the material is left 20 min at stress zero to equilibrate both thermally and mechanically. At the end of this equilibration time, the position $\varphi$ corresponding to zero stress is defined as the equilibrium position $\gamma(\varphi)=0$. Stress relaxation tests are performed over the duration of two hours.

*2.3 Methods for master curve construction, stress relaxation modulus conversion and mechanical spectrum analysis*

The SAOS data can be collapsed onto a single master curve. To check the applicability of the rescaling, we first consider the van Gurp-Palmen plot (Fig. S4), where the phase angle, $\delta$ is plotted as a function of the complex modulus, $G^*$. In the case the data can be coherently rescaled, this graph produces a single curve without introducing user-defined shift factors. Then the shift factors $\alpha_i$ are introduced to build the graph $\delta(\alpha_i \omega)$ (Fig.S5) in such a way that they reproduce the shape of the van Gurp-Palmen plot. The vertical factors $\beta_i$ are taken as inversely proportional to the temperature in the rubbery regime, to account for entropic elasticity, and as constant in the glassy regime.

To extend the frequency domain, conversion of the time dependent relaxation modulus $G(t)$ into the angular frequency dependent complex modulus $G^*(\omega)$ is performed. This conversion formally corresponds to a Laplace Transform of the relaxation modulus $G(t)$, convoluted with the experimental window. This is known to be a mathematically ill posed problem [23], in other words the solution is not unique. Different methods to perform the inversion have been proposed [24,25]. Here we use a custom algorithm, inspired from an algorithm developed for analogous problems [26], structured as follows:

- We consider a hypothetical distribution of relaxation frequencies

$$H^h(\omega) = \Sigma_i H(\Omega_i)\delta_D(\omega - \Omega_i) \qquad (1)$$

where $\delta_D$ is Dirac's delta function and $H(\Omega_i)$ is the amplitude of the mode associated to the relaxation frequency $\Omega_i$.

- The relaxation modulus $G^h(t)$ is computed from $H^h(\omega)$:

$$G^h(t) = \Sigma_i \frac{H(\Omega_i)}{\Omega_i} e^{-t\Omega_i} \qquad (2)$$

- The difference between the calculated and experimental complex moduli is measured as:

$$\Delta = \int dt \frac{|G^e(t) - G^h(t)|}{G^e(t)} \qquad (3)$$

where the superscript *h* stands for calculated from the hypothesis, and *e* stands for experimental.

- The amplitudes of the relaxation components $H_i$ are then optimized using a built-in Matlab algorithm [26] to minimize the difference $\Delta$ between the modeled and the experimental moduli. The optimization is done either by assuming a certain functional form for $H(\omega)$ – procedure used in section 3.3, or letting free all the amplitudes of the modes $H_i(\Omega_i)$ – procedure used in section 3.4.

Once optimization is performed, the distribution of relaxation frequencies that best fits the experimental data is noted $H^m(\omega)$ and called model relaxation spectrum.

- The spectrum $H^m(\omega)$ thus obtained has an intrinsic interest, as discussed below in sections 3.4 and 3.5. It can be used likewise to predict the relaxation modulus decay $G(t)$ using formula (2) and the values of storage $G'$ and loss $G''$ moduli as a function of the angular frequency using formulas (4) and (5).

$$G'(\omega) = \Sigma_i H(\Omega_i) \frac{(\omega/\Omega_i)^2}{1 + (\omega/\Omega_i)^2} \qquad (4) \qquad G''(\omega) = \Sigma_i H(\Omega_i) \frac{\omega/\Omega_i}{1 + (\omega/\Omega_i)^2} \qquad (5)$$

Overlaps of such predicted $G(t)$, $G'(\omega)$ and $G''(\omega)$ with experimental ones are shown in sections 3.3 and 3.4.

## 3. Results
*3.1 Time stability of stress relaxation properties*

Previous studies performed at 200°C on the same material showed a gradual and irreversible increase of the elastic modulus [20]. This has been attributed to the presence of secondary reactions conducive to the formation of non-exchangeable ether links. In the same work, the viscoelastic moduli (at 1 rad/s) were observed to be constant for a reasonable amount of time (7 h) at the temperature of 180°C. To better investigate the evolution of the mechanical properties, and define the safe operational temperature window, 2 hours-long stress relaxation tests have been repeated up to 7 times. The results are presented in Figure 2. Throughout these repeated experiments at 180°C, the elastic modulus and the relaxation time do not show systematic evolution, proving that the material is stable within this time and temperature window.

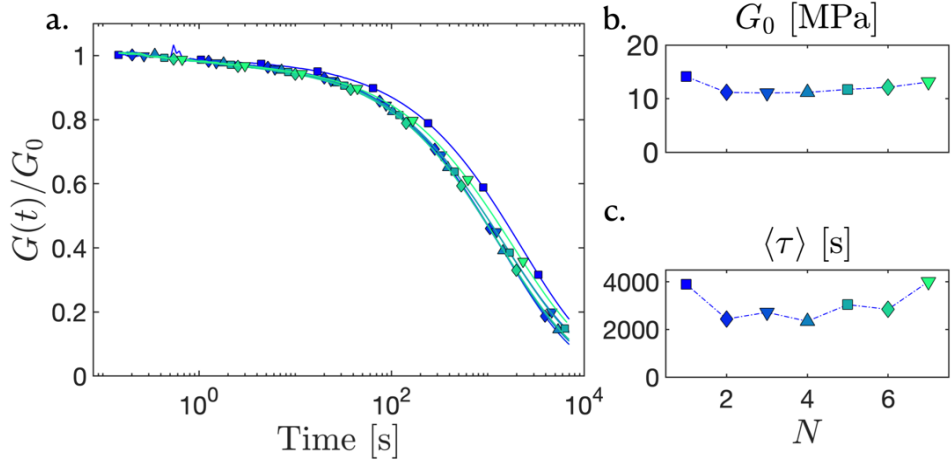

**Fig. 2.** Repeated stress relaxation tests at 180 °C. The thermal stability of the system is crucial for reprocessability; stress relaxation plots show that the material is able to relax the applied stress completely several times with minor change in its mechanical properties: **a.** Time dependence of the normalized modulus $G(t)/G_0$ during repeated tests. The data are fitted with a stretched exponential $G(t) = G_0 \exp(-(t/\tau)^s)$ **b.** Value of the elastic modulus $G_0$ at the beginning of the $N^{th}$ test **c.** Average relaxation time $\langle \tau \rangle = \frac{\tau}{s}\Gamma\left(\frac{1}{s}\right)$, as measured during the $N^{th}$ test.

*3.2 SAOS*

To characterize the mechanical response in a range of temperature encompassing the ambient (or slightly above) as well as temperatures well above the glass transition, where reprocessing can be envisioned, we hereby report small amplitude oscillatory shear (SAOS) experiments performed between 30°C and 180°C. At first, we determine the linear regime at 1 Hz for three temperatures, below $T_\alpha$, in the vicinity of $T_\alpha$, and above $T_\alpha$. At 40°C ($T < T_\alpha$) the material shows a storage shear modulus of the order of $G' \approx 1$ GPa and a linear range of deformation of about $\gamma \approx 0.02\%$. At 90°C ($T \approx T_\alpha$), the elasticity drops to a value of $G' \approx 10$ MPa, while the linear deformation range increases to $\gamma > 2\%$.

The linear regime being established, SAOS experiments are performed at angular frequency ranging from $\omega = 0.01$ to 100 rad/s. Frequency sweep tests are performed at different temperatures, from 30 to 180 °C, with steps of approximately 10 degrees. Figures 3a-d show typical viscoelastic spectra measured when the temperature is varied. The values of the storage modulus $G'$ and the loss factor $\tan \delta = G''/G'$ at 1 Hz are represented in Figs. 3e–f as a function of temperature. In the lower temperature range, the trace of $G'$ (Fig. 3e) shows a plateau, the glassy regime. Around $T = 80$°C, the value of $G'$ drops (Fig. 3e) whereas the loss factor $\tan \delta$ reaches its maximum value (Fig. 3f) featuring the $\alpha$ relaxation. Above 80°C a second plateau typical of the rubbery regime in a crosslinked material is visible. In Fig. 3g we show the whole data set of frequency sweeps taken at the different temperatures, butted together into a single master-curve. Reaching this type of time-temperature superposition requires the application of temperature dependent vertical factors $\beta(T)$ and horizontal factors $\alpha(T)$, whose meaning will be discussed hereafter. The $\beta$ factors follow an inverse dependence on the temperature to account for the change of entropic elasticity with temperature [17]. The $\alpha$ factors follow a sigmoïdal dependance against inverse temperature. The data above the glass transition temperature can be equally well fitted to the Williams-Landel-Ferry equation (blue continuous line in Fig. 3i), or to two distinct Arrhenius equations (red and magenta dotted lines). The activation energies of the two processes are for the glassy part $E_G = 680 \pm 50$ kJ/mol, and for the vitrimeric part $E_V = 130 \pm 10$ kJ/mol, compatible with values in the literature [20]. As we discuss in the conclusion, we favor here the double Arrhenius description as it is more suited for this system where two distinct physical processes (glassy and vitrimeric) are expected.

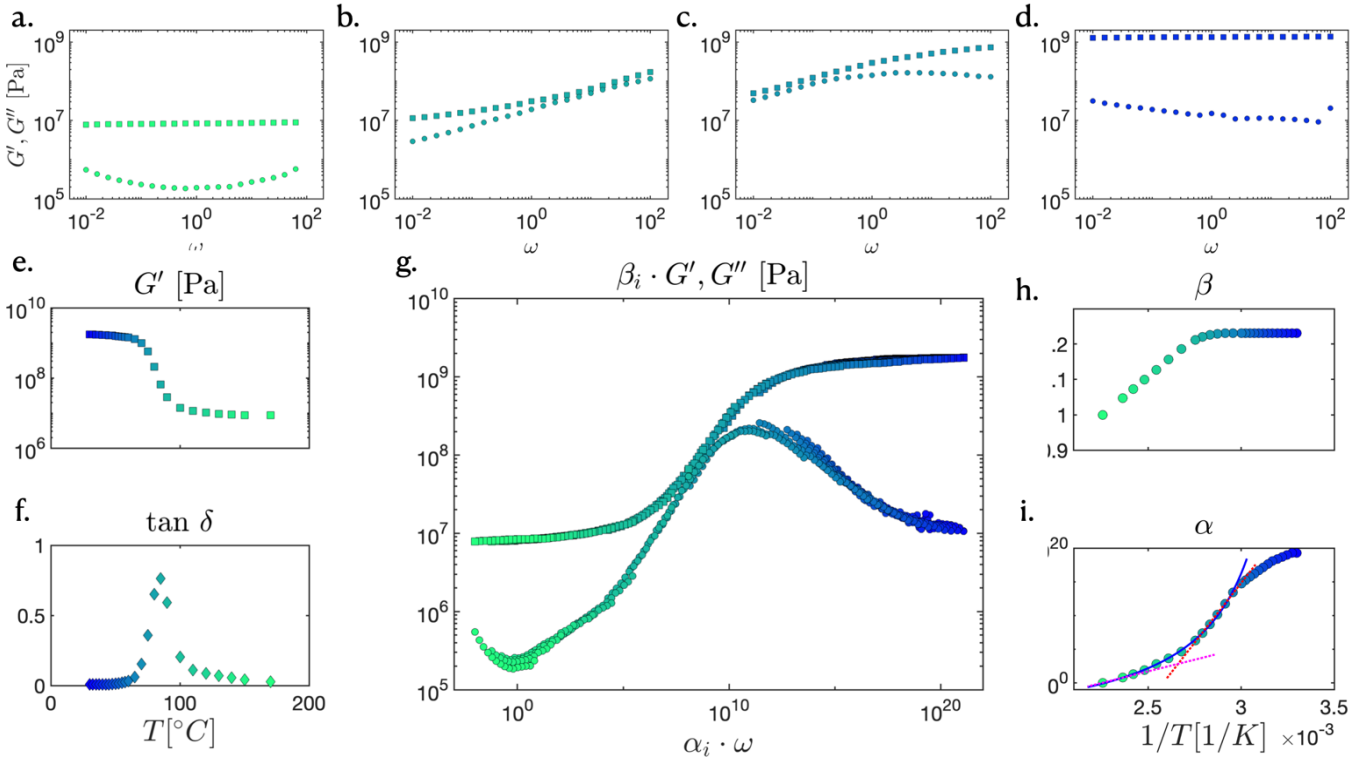

**Fig. 3.** SAOS – **a-d** Raw data from SAOS tests performed at T = 180 °C, 85°C, 75°C, 30 °C, from right to left in the range of angular frequencies [0.01–100 rad/s]. **e, f** Temperature dependence of the shear elastic modulus, $G'$ and the loss factor, tan $\delta$ at 1 Hz. **g.** Time temperature superposition of SAOS data (**a-d**). **h.** Vertical shift factors $\beta_i$ (1/T) applied to account for the variation of the entropic elasticity effect with temperature. **i.** Horizontal shift factors $\alpha_i$ (1/T), above the glass transition temperature the data are well fitted to two distinct Arrhenius functions, revealing two different activation energies, one associated with the glass transition (red dotted line) and the other one with the covalent bond exchange dynamics (magenta dotted line).

The master-curve shown in Fig. 3g, obtained by applying time temperature superposition, encompasses an impressive range of frequencies. One can recognize the glassy plateau, the glass transition regime, the intermediate regime where the moduli are nearly superimposed and vary as $\omega^{1/2}$ reminiscent of Rouse relaxation, the rubbery plateau, and an upturn of the viscous modulus at the lower frequencies, signaling the onset of the terminal relaxation process observed in stress relaxation (Fig. 2a).

*3.3 Conversion of time-dependent creep and stress relaxation data to frequency dependent complex moduli.*

To extend the accessed frequency range beyond the limits of the SAOS experiments, additional data can be extracted from stress relaxation and creep experiments [24, 25, 28]. In the linear regime, the creep compliance *J(t)* and the relaxation modulus *G(t)* can be both converted to frequency-dependent viscoelastic moduli and compared to the results of direct SAOS experiments. These methodologies are particularly well-suited to access low frequencies. The minimum accessed frequency corresponds to the inverse of the experiment duration. The linear creep experiment shown in Figure 4a exhibits a power-law flow regime where the creep compliance can be fitted to the function $J(t) = a + b\, t^a$. The initial oscillations are due to a phenomenon known as the creep-ringing effect, which arises from the coupling between inertia and elasticity [27], and are discarded in our analysis. The linear creep compliance is then numerically converted into viscoelastic moduli through the Schwarzl formula [28]. Since the creep compliance deviates from the Maxwell-like response $J(t) = a + b\, t$, the typical variations $G'' \approx \omega$, and $G' \approx \omega^2$ expected in the terminal relaxation regime are not observed. Yet, the crossover point, where $G' = G''$ is well detected (Fig. 4b).

Stress relaxation experiments give qualitatively similar results, in particular complete relaxation of stress is observed at long enough times (Fig. 4c). As introduced in the methodological section 2.2, the stress relaxation modulus has been converted into viscoelastic moduli passing through the determination of the model spectrum $H(\omega)$. The fact that the relaxation modulus is not a simple exponential indicates that the relaxation mechanism is heterogeneous, and can be characterized statistically by a broad distribution of relaxation times, $H(\omega)$, containing more than one component. In this section we follow the methodology described in Section 2.2, searching for the mechanical spectrum $H^m(\omega)$ that best matches the experimental data within a family of functions with the same functional form (e.g. either gaussian, or log-normal, or power-law or others). Different functional forms have been tested and the one producing the best result is a power-law distribution with an exponential cut-off:

$$H(\omega) = \frac{A}{\omega^\epsilon} \exp(-\omega_0/\omega) \tag{6}$$

where $\epsilon = 1.76$, $A = 1.9\ 10^4$, $\omega_0 = 2.9\ 10^{-4}$ rad/s. This corresponds to an extremely broad distribution of relaxation frequencies (FWHM 2 decades). This procedure simultaneously provides us with an approximated modulus $G(t)$ represented by the red line in Figure 4c, the mechanical spectrum $H^m(\omega)$ (using Eq 5), and the viscoelastic moduli (using Eqs 3 and 4) shown by red lines in Figure 4d and compared to the direct oscillatory measurements.

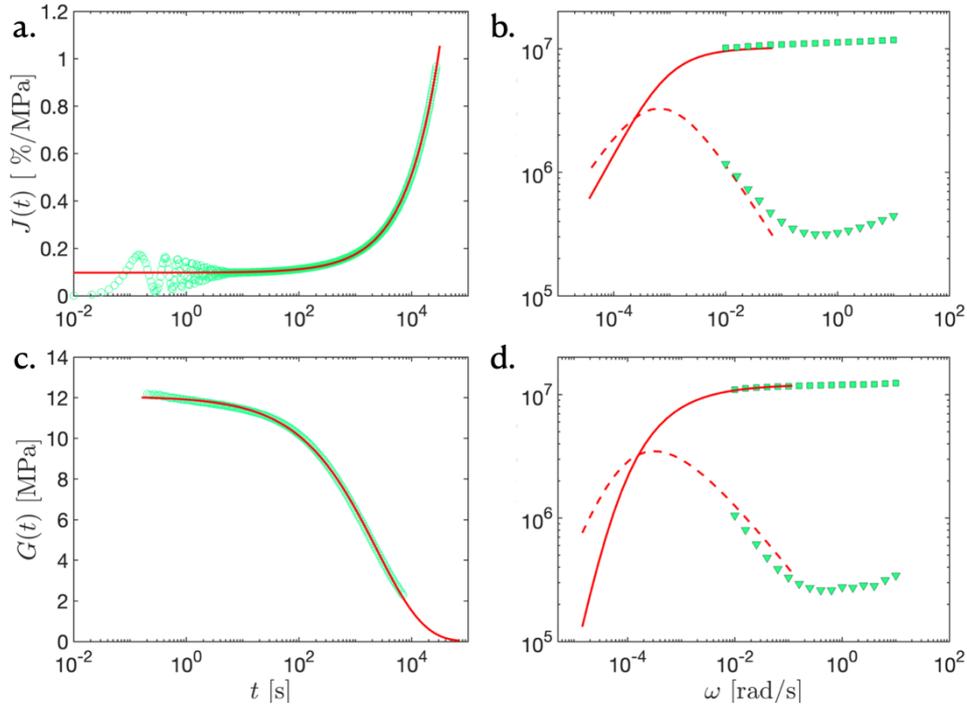

**Fig. 4.** Creep (top) and stress relaxation (bottom) - **a.** Creep compliance $J(t)$ at 180°C and fit to the power law $J(t) = J_0 + k\, t^\varepsilon$ (red line). **b.** Conversion of the creep compliance into viscoelastic moduli following the Schwarlz method (red lines) compared with direct measurements of viscoelastic moduli by SAOS at 180°C (symbols). **c.** Stress relaxation modulus $G(t)$ at 180°C (symbols) and fit (red line) to a superposition of Maxwell modes $G(t) = \Sigma_i G_i \exp(-t/\tau_i)$. **d.** Analytical conversion of the fitted data into viscoelastic moduli (red lines) compared with direct measurements of viscoelastic moduli by SAOS (symbols).

## 3.4 Master Curve, Mechanical Spectrum

The agreement between the different experimental protocols we implemented allows us to build a unique master curve giving the material response from room temperature, where it is dominated by the glassy behavior, to high temperatures where the dynamic bond exchanges prevail leading to terminal relaxation (Fig. 5a). The complete master curve is further analyzed using the routine introduced earlier in Section 2.3 to extract the relaxation spectrum, $H^m(\omega)$ on the whole frequency window. In this case, instead of optimizing the spectrum within a family of functions with given form, we let each component $H_i$ of the spectrum free, the number of degrees of freedom being chosen to be 40 and logarithmically spaced in the interval $\Omega \in [10^{-7}$–$10^{22}]$ rad/s.

The optimization procedure (see section 2.3) produces the model spectrum $H^m(\omega)$, which is plotted in Figure 5b. Two distinct families of modes appear clearly, which will be hereafter recognized as the *glassy modes* (in bright green), and the *vitrimeric modes* (in dark green). In the model, each of the two contributions is described as a log-normal distribution:

$$H_G = H_G^0 \exp\left(-\frac{(\log \omega - \log \omega_G)^2}{S_G}\right) \quad ; \quad H_V = H_V^0 \exp\left(-\frac{(\log \omega - \log \omega_V)^2}{S_V}\right) \tag{7}$$

where $\omega_G$ (resp. $\omega_V$) is the central frequency of the distribution for the glassy (resp. vitrimeric) modes, $H_G^0$ and $H_V^0$ are their respective amplitudes, and $S_G$, $S_V$ are the parameters describing the width of the distributions.

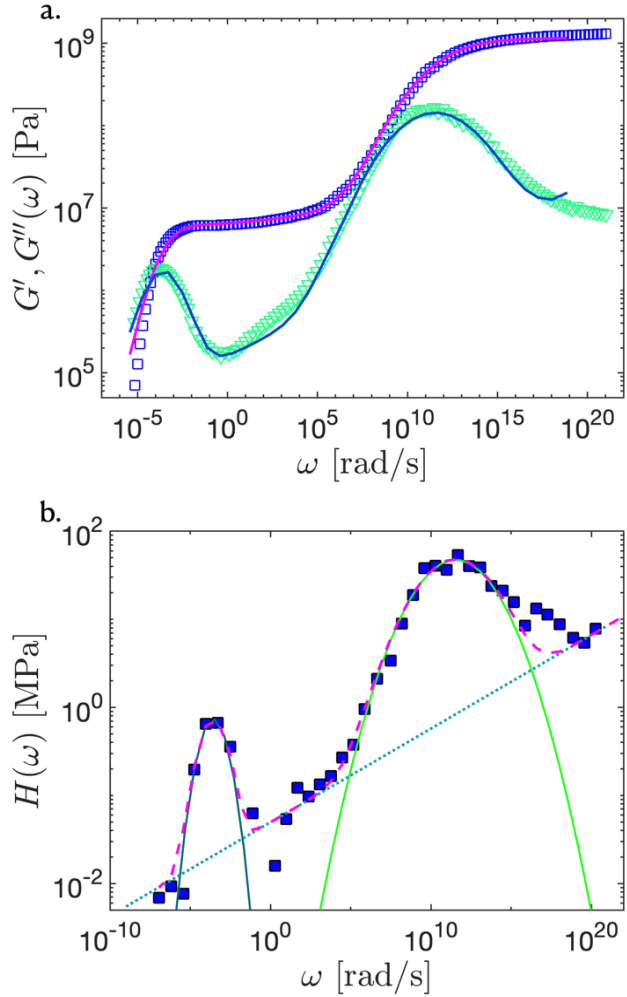

**Fig. 5. a.** Master curve (symbols) obtained combining SAOS (Fig. 3g) and stress relaxation tests (Fig. 4d), and fit (magenta and blue lines) to a superposition of Maxwell modes (Eq 5). **b.** Distribution of relaxation times $H^m(\omega)$ (squares) minimizing the difference between modeled and experimental data (Eq 2), and result of the fit to a three term model (magenta dashed line): the log normal distribution $H_1(\omega)$ describing the glassy modes (bright green continuous line), the log normal distribution $H_2(\omega)$ describing the vitrimeric modes (dark green continuous line) and a phenomenological power law term (dotted line).

To these two distributions of modes with clear physical meaning, we add a phenomenological power law term (dotted line in Fig. 5b). This term accounts for additional modes not captured by the bimodal distribution and responsible for finite dissipation in the range between the vitrimeric and glassy modes (frequency $\omega \in [1, 10^5]$ rad/s), as well as above the glass transition ($\omega > 10^{17}$ rad/s). From the fit of the spectrum (dashed magenta line) we can transform this back to the viscoelastic moduli following Eq (3) (magenta and blue lines in Fig. 5a) providing an excellent approximation of the data, including the shape of the two local maxima of $G''$.

The two contributions have different physical origin and different activation energies, as observed from the horizontal shift factors $\alpha(T)$ (Fig. 3e). The spectrum of relaxation times can be approximated by a sum of two contributions each with a specific temperature dependence:

$$H(\omega, T) = H_G(\omega, T) + H_V(\omega, T) \tag{8}$$

We propose here to distinguish the concept of time-temperature equivalence from that of time temperature superposition. The consequence of the presence of two different processes is that the effect of time is not **equivalent** to that of temperature, in this sense the master cure represented in Figure 5, does not correspond to the response of the material at any given temperature over an extended range of frequencies. Nonetheless, the two processes are sufficiently well separated that in each experiment (frequency sweep at a given temperature) they are never observed together, and this allows to consistently **superimpose** over a limited frequency domain the data acquired at different temperatures, with the objective of producing the master curve. We have been able to identify the two dominant relaxation processes with their respective temperature-dependence leading to relaxation frequencies that depend on temperature according to Arrhenius-like equations:

$$\omega_G = \omega_G^0 \exp\left(-\frac{E_G}{RT}\right) \quad ; \quad \omega_V = \omega_V^0 \exp\left(-\frac{E_V}{RT}\right) \tag{9}$$

where the activation energies are experimentally identified (Fig. 3i). As a result, when the temperature is changed the two modes change in a different way, and in particular their distance increases when the temperature increases (Fig. 6a).

Modeling the actual temperature evolution of the mechanic spectrum $\widetilde{H}(\omega, T)$, we can compute the viscoelastic moduli $\widetilde{G}', \widetilde{G}''(\omega, T)$ at any temperature (Fig. 6b), where we use the symbol tilde to explicit that these quantities are not directly measurable and they have been computed with the method that we just described. At T=180 °C they resemble to the master curve in Fig. 5a, in particular we distinguish the two plateaus, the terminal relaxation, and the intermediate regime where the moduli vary like $\omega^{1/2}$. As the temperature is decreased, the extension of the plateau is reduced, as a consequence of the difference in the activation energies. At room temperature, the moduli that we predict have no rubbery plateau, as the lifetime of covalent bonds is close to that of glassy modes, and so only one relaxation is observed.

The master curve (Fig 5a) shows us at a glance the mechanical behavior in the relevant temperature regime, with the *caveat* that we have discussed about the difference between time temperature superposition (allowing to build a master curve) and time temperature equivalence (giving to the curve the meaning of mechanical response at any temperature over the rescaled frequency regime*)*.

Extracting the spectrum, analyzing it in terms of physical processes, and finally recombining the two contributions with the adequate temperature dependence allow us to predict the viscoelastic moduli over a broad range of temperatures and frequencies. Despite the failure of time temperature equivalence, the proposed analysis allows us to take advantage of the speeding-up effect of the temperature to access range of frequencies that would be otherwise inaccessible.

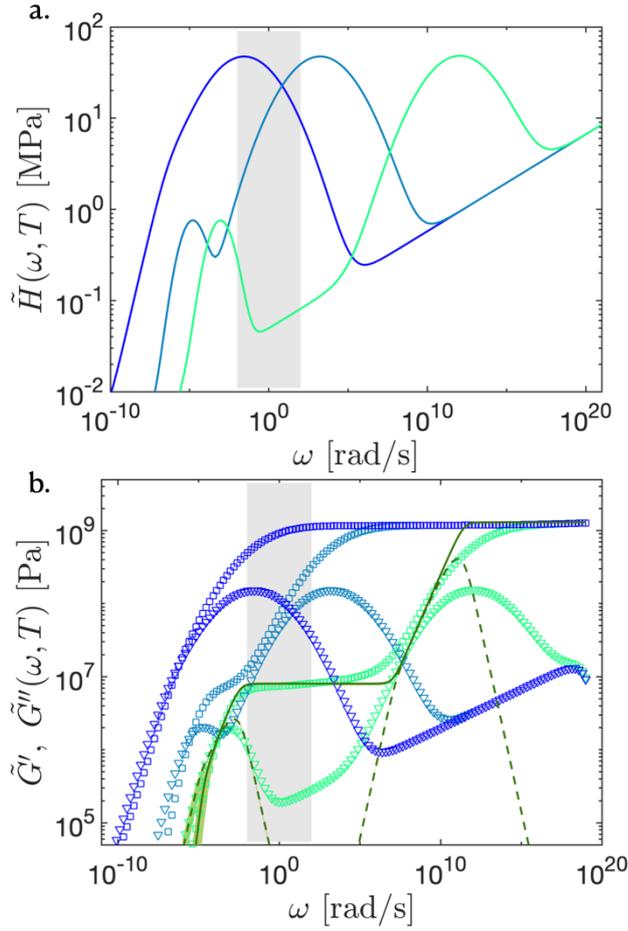

**Fig. 6.** a) Mechanical spectrum $H(\omega)$ computed at different temperatures (electric blue 25 °C, blue 80 °C, bright green 180 °C), implementing for each family of relaxation modes (glassy and vitrimeric) its own temperature dependence. b) Computed viscoelastic moduli obtained by transforming analytically the spectrum $H(\omega)$.

### 3.5 Topology of the network

Our objective here is to model the viscoelastic properties we have obtained in terms of the sticky Rouse model. The interest is not to propose a faithful modeling of the system, but to show in an example how modeling of the material response allows to access relevant microscopic parameters. In the simplified sticky Rouse model [29, 21] the material is represented through a *linear* chain composed of $N$ monomers. A part of this monomers has a certain relaxation time $\tau$, and another part of the monomers ($N_X$) has a slower relaxation time ($\tau_X$). The slower monomers represent the dynamic links between the different chains that compose the network. The fingerprint of this model is the transition where the moduli $G'$ and $G''$ are parallel with a one-half exponent power law behavior, both near the glass transition and at the beginning of the vitrimeric relaxation.

Considering the response $\tilde{G}(\omega)$ at 180 ºC (Fig 6b), the physical parameters that we can extract are: the ratio between the two plateaus providing $N/N_x$ (in this case $\simeq 150$), meaning the ratio between free monomers and dynamically crosslinked ones; the extension of the second Rouse regime (before the beginning of the terminal relaxation with 1 and 2 exponents) giving the number of dynamic crosslinks permanently connected $N_x$ (in this case $\simeq 14 \pm 6$).

The main simplification made in this model is to consider that the network is a dynamic assemble of linear chains, neglecting any branching of the permanently connected structures (see SI for some additional details on the modeling). A detailed topological analysis starting from the basic macromolecules (Fig. 1) which would predict the topological characteristic of the network, goes beyond the scope of this work and will be studied in a following paper [30]. For the moment we only anticipate that the task to predict the network topology, and in particular the number of permanently connected dynamic crosslinks, is not trivial and has a great impact on the mechanical properties of the network, and that an approach that combines solid-state and fluid-state rheology and modeling is promising to access this kind of information and rationalize the response of the system on physical bases.

## 4. Conclusions

In the first part of the paper we investigate the solid state and fluid state viscoelastic properties of a recently developed high $T_g$ vitrimer [20Errore. Il segnalibro non è definito.] using torsional rheology. We obtain the variations of the elastic modulus as function of the temperature and we determine the glass transition temperature. We also demonstrate that the material exhibits terminal relaxation and is indeed a vitrimeric material. The mechanical properties (in particular the capability to relax the stress) are thermally stable at processing temperatures, and we obtain the shift factors as function of the temperature and two activation energies.

In the second part of the paper, we combine this information to discuss the applicability of TTS. We show that we can collapse the data onto a single master curve, at the same time we identify two different relaxation mechanisms (glassy and vitrimeric). The shift factors can be either described by a WLF or by a double Arrhenius. What is the more appropriate approach between the two is actually debated in the literature, and the correct answer is expected to be system dependent. In our case, in terms of temperature the glassy regime is relatively close to the vitrimeric one, so it is not possible to say if the deviations from Arrhenius are only due to an interplay with the glassy dynamics, or to the fact that we only capture the beginning of the vitrimeric regime. It is anyhow evident that two processes are present, and so it seems more meaningful to describe the shift factors with two distinct activation energies.

The presence of different mechanism with different temperature dependence makes the material thermo-rheologically complex, and thus time temperature equivalence fails. We can explain the fact that nonetheless the data collapse in a single master curve as follows. Indeed the two relaxation processes are well enough separated (in frequencies), as compared to our measuring frequency window, and thus in our experiments we never detect the two processes at the same time: we detect either the glassy modes, or the rubbery plateau (no relaxation), or the terminal vitrimeric relaxation. This observation reconciles the contradiction found in literature and presented in the introduction.

The shift factors $\alpha(T)$, give us the temperature dependence of the two processes; the master curve gives us the distribution of relaxation modes $H^m(\omega)$, from which we separate and characterize the two families of relaxation modes. With these two elements we can recombine the two modes each one with its own temperature dependence to obtain the mechanical response at any temperature.

This methodology, based on standard tools of linear rheology, extends the state-of-the-art rheological practice that in case of thermo-rheological complexity does not provide a modeling of the temperature dependent response over a wide frequency range.

Finally, the viscoelastic moduli (Fig. 6b) are compared with the simplified sticky Rouse model that gives access to some relevant topological parameters of the network.

To summarize, we stress at first the importance of rheological measurement also in the solid regime, to obtain a complete characterization of the relaxation modes originating the mechanical response of the system. These measurements are made possible thanks in torsional rheology. Secondly, we stress the fact that the methodology of measurements and analysis that we presented in this paper opens the path to compare the mechanical response to microscopic models.

## 6. Acknowledgements

We acknowledge helpful discussion in the framework of the VITRIMAT project. PE thanks the CNRS for financial support. Anton Paar France SAS is gratefully acknowledged for loan of the MCR702 rheometer. We acknowledge funding from the ANR through the MATVIT project (ANR-18-CE06-0026-01). We thank Rémi Fournier for preliminary experiments using torsional rheology.